\documentclass[RNAAS]{aastex631}

\submitjournal{RNAAS}
\usepackage{amsmath}
\usepackage{booktabs}
\usepackage{float}
\usepackage{longtable}

\shorttitle{New ZTF CPVS Classification}
\shortauthors{Cheung et al.}
\graphicspath{{./}{figures/}}

\begin{document}

\title{A New Classification Model for the ZTF Catalog of Periodic Variable Stars}

\author[0000-0002-0814-3378]{Siu-Hei Cheung}
\affiliation{Department of Physics and Institute of Theoretical Physics, The Chinese University of Hong Kong, Shatin, N.T., Hong Kong}
\affiliation{Center for Computational Astrophysics, Flatiron Institute, New York, NY 10010, USA}
\author[0000-0002-5814-4061]{V. Ashley Villar}
\affiliation{Department of Astronomy and Astrophysics; Institute for Computational and Data Sciences; Institute for Gravitation and the Cosmos. The Pennsylvania State University University Park, PA, USA}
\author[0000-0003-2776-082X]{Ho-Sang Chan}
\affiliation{Department of Physics and Institute of Theoretical Physics, The Chinese University of Hong Kong, Shatin, N.T., Hong Kong}
\affiliation{Center for Computational Astrophysics, Flatiron Institute, New York, NY 10010, USA}
\author{Shirley Ho}
\affiliation{Center for Computational Astrophysics, Flatiron Institute, New York, NY 10010, USA; New York University, New York, NY 10011; Princeton University, Princeton, NJ 08540; Carnegie Mellon University, Pittsburgh, PA 15213}

\begin{abstract}
Using the second data release from the Zwicky Transient Facility \citep[ZTF,][]{2019PASP..131a8002B}, \citet{Chen_2020} created a ZTF Catalog of Periodic Variable Stars (ZTF CPVS) of $781,602$ periodic variables stars (PVSs) with $11$ class labels. Here, we provide a new classification model of PVSs in the ZTF CPVS using a convolutional variational autoencoder and hierarchical random forest. We cross-match the sky-coordinate of PVSs in the ZTF CPVS with those presented in the SIMBAD catalog. We identify non-stellar objects that are not previously classified, including extragalactic objects such as Quasi-Stellar Objects, Active Galactic Nuclei, supernovae and planetary nebulae. We then create a new labelled training set with $13$ classes in two levels. We obtain a reasonable level of completeness ($\gtrsim 90$ \%) for certain classes of PVSs, although we have poorer completeness in other classes ($\sim40$ \% in some cases). Our new labels for the ZTF CPVS are available via Zenodo \citet{cheung_siu_hei_2021_5764899}.
\end{abstract}


\section{The ZTF CPVS} \label{sec:intro}
The ZTF is a wide-field, optical survey conducted using a $48$-inch Schmidt telescope with a $47$ deg$^2$ field of view \citep{2019PASP..131a8002B}. \citet{Chen_2020} made use of the second data release of the ZTF to create the ZTF CPVS. They search for and classify new PVSs down to an $r$-band magnitude of $\sim 20.6$. By measuring the $g$- and $r$-band periods, phase difference, amplitudes, absolute Wesenheit magnitude and adjusted $R^{2}$ (which represents how well data are fitted by the Fourier function), they are able to group PVSs into $11$ distinct types using linear cuts of these observational features. Moreover, $79.5\%$ of the PVSs presented in the ZTF CPVS are newly classified objects. They report a misclassification rate of $2\%$ when compared to other photometrically classified samples, such as  ATLAS \citep{2018AJ....156..241H}, WISE \citep{2018ApJS..237...28C}, ASAS-AN \citep{2018MNRAS.477.3145J}, and the CATALINA \citep{Drake_2014, 2017MNRAS.469.3688D} catalogs. \\

Here we present a new set of classifications based on a deep generative feature space and an independent set of class labels obtained from the SIMBAD catalog \citep{2000A&AS..143....9W}.

\section{Methodology}
Our classifier is based on learned latent features generated in Chan et al. 2021 (in prep, hereafter C21), as well as hand-engineered features from the ZTF CPVS, including the periods, amplitudes and mean magnitudes of both the $g$- and $r$-band light curves. The C21 latent features are generated by a convolutional variational autoencoder, with ten features describing the light curves of each object. We extract object labels from the SIMBAD catalog \citep{2000A&AS..143....9W} by cross-matching the sky coordinates of PVSs with those in the ZTF CPVS \citep[using \texttt{Astroquery},][]{2019AJ....157...98G}. We find $31,541$ successfully cross-matched objects which we use as the training set. We then construct 13 class labels in two levels. The first level contains Active Galactic Nuclei-like objects (AGNL, including blazars and quasars), cepheids (CEP), eclipsing binaries (EB), long-period variables (LPV), Mira variables (Mira), RR Lyraes (RR), and the catch-all categories of other pulsating variables (Pul$_\mathrm{oth}$), and peculiar types (Pec). The second level is a further classification of the Pec class. They include carbon stars (C-Type), horizontal branch stars (HB), red giant branch stars (RGB), S-Type stars (S-Type), young stellar object-like (YSOL), and other variables (V$_\mathrm{oth}$). We split the data set into a training-to-test set ratio of $7:3$ by using python package \texttt{scikit-learn} \citep{scikit-learn}. We note that our training set is highly imbalanced, with the largest class containing $10,745$ objects and the smallest containing just $41$ objects. We balance the training set using the python package \texttt{imbalanced-learn} \citep{JMLR:v18:16-365} with default learning parameters, which performs synthetic minority resampling \citep{chawla2002smote, JMLR:v18:16-365}. Finally, the hand-engineered features and the latent vectors $\vec{\mu}$ of the cross-matched objects are fed into the hierarchical random forest provided by \texttt{imbalanced-learn} for training, with no hyper-parameter optimization conducted.

\section{Results}
\begin{figure*}[ht!] 
	\centering
	\includegraphics[width=1.0\linewidth]{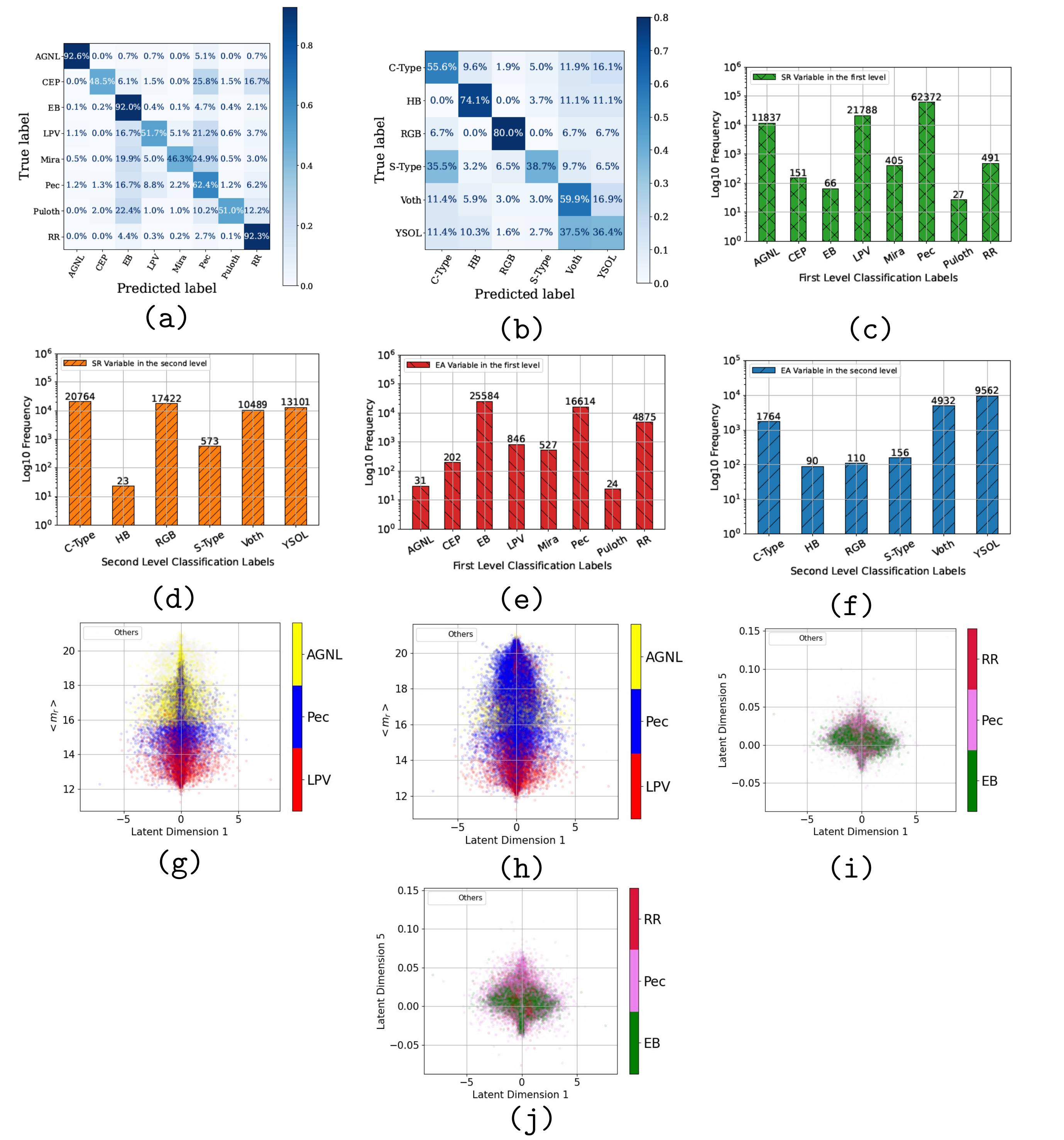}
	\caption{(a) Confusion matrix of the first level hierarchical random forest for the test set. We show in each row the percentage of each class label being predicted as a particular class. (b) Same as (a), but for the second level. (c) Class break-down for the SR variables in the ZTF CPVS with respect to the first level of our new classification model. (d) Same as (c), but for the second level. (e) Same as (c), but for the EA variables. (f) Same as (d), but for the EA variables. (g), (h), (i), and (j) Illustrations of the latent representations used in our training of the hierarchical random forest. In (g), we plot the distribution of AGNL, LPV, and Pec variables that are previously classified as SR in the ZTF CPVS. (h) is the same as (g), but for the distribution of all AGNL, LPV, and Pec variables. (i) and (j) are the same as (g) and (h), respectively. In (i), we plot the distribution of EB, Pec, and RR variables that are previously classified as EA in the ZTF CPVS. (j) is the same as (i), but for the distribution of all EB, Pec, and RR variables.\label{fig:results}}
\end{figure*}
Here we discuss the classification results and their implications. We show the confusion matrix of our classification results for the test set in Figure \ref{fig:results} (a) and (b). Our new classification model has excellent classification completeness for certain classes of objects, such as AGNL ($93$ \%), RR ($92$ \%), EB ($92$ \%). However, the completeness for some objects, such as CEP ($48$ \%), Mira ($46$ \%), and YSOL ($36$ \%) are poorer. This is likely due to insufficient samples of these classes in our training set. In addition, we compute the class-averaged precision and accuracy. In addition, we obtained a class-averaged accuracy of $0.97$ ($0.84$), precision of $0.71$ ($0.45$), error rate of $0.03$ ($0.16$), f1-score of $0.69$ ($0.50$), and purity of $0.87$ ($0.54$) in the first (second) level of our new classification. The second level classification performs worse than the first level counterpart, which may also be due to insufficient samples presented in the data set or intrinsic overlap in labels (e.g., C-Type and S-Type variables may be intrinsically very similar). \\

We highlight specific classifications which are distinct from the ZTF CPVS. In particular, we find that the SR variable category in the ZTF CPVS very likely consists of multiple classes. As shown in Figure \ref{fig:results} (c) and (d), this class includes LPV, AGNL, C-Type, RGB, YSOL, and V$_\mathrm{oth}$ variables based on our classifier. Furthermore, we note that the ZTF CPVS provides class labels only for \textit{galactic} periodic variable stars. However, our cross-matching and classification results reveal that the ZTF CPVS may contain \textit{non-stellar} variables. For instance, we find from our cross-matched results planetary nebulae, supernovae and active galactic nuclei; Furthermore, $11,837$ of the SR variables in the ZTF CPVS are classified as AGNL with our classifier. We hope that a closer comparison of these labels can lead to improved purity in the ZTF CPVS.

\section{Conclusion}
We present a new classifier and photometric labels for PVSs in the ZTF CPVS. Our new classifier is a 2-layered hierarchical random forest that uses latent features generated by a convolutional variational autoencoder and class labels given from the SIMBAD catalog. We obtain a reasonable level of completeness ($\gtrsim 90$ \%) for certain classes of PVSs, although we have poorer completeness in other classes ($\sim40$ \% in some cases).  Furthermore, we find non-stellar and extra-galactic objects within the ZTF CPVS which were not previously identified. Finally, our classifications are available on Zenodo. 

\newpage
\bibliography{sample631}{}
\bibliographystyle{aasjournal}



\end{document}